%% file: radcor00.tex
\documentclass[12pt]{article}
\usepackage{epsfig}

\newcommand\pubnumber{}
\newcommand\pubdate{\today}
\newcommand\hepnumber{hep-ph/0101135}

\def\csumb{Theoretical Physics Dept.\\
Fermi National Accelerator Laboratory, Batavia, IL 60510 USA}

\def\Title#1{\begin{center} {\Large\bf #1 } \end{center}}
\def\Author#1{\begin{center}{ \sc #1} \end{center}}
\def\Address#1{\begin{center}{ \it #1} \end{center}}

\newcommand\pubblock{\rightline{\begin{tabular}{l} \pubnumber\\
         \pubdate\\ \hepnumber \end{tabular}}}
\newenvironment{Abstract}{\begin{quotation}  }{\end{quotation}}
\newenvironment{Presented}{\begin{quotation} \begin{center} 
             Presented at the\end{center}
      \begin{center}\begin{large}}{\end{large}\end{center} \end{quotation}}
\def\Acknowledgments{\bigskip  \bigskip \begin{center}
          \large\bf Acknowledgments\end{center}}

\makeatletter
\def\section{\@startsection{section}{0}{\z@}{5.5ex plus .5ex minus
 1.5ex}{2.3ex plus .2ex}{\large\bf}}
\def\subsection{\@startsection{subsection}{1}{\z@}{3.5ex plus .5ex minus
 1.5ex}{1.3ex plus .2ex}{\normalsize\bf}}
\def\subsubsection{\@startsection{subsubsection}{2}{\z@}{-3.5ex plus
-1ex minus  -.2ex}{2.3ex plus .2ex}{\normalsize\sl}}

\renewcommand{\@makecaption}[2]{%
   \vskip 10pt
   \setbox\@tempboxa\hbox{\small #1: #2}
   \ifdim \wd\@tempboxa >\hsize     
       \small #1: #2\par          
     \else                        
       \hbox to\hsize{\hfil\box\@tempboxa\hfil}
   \fi}

 \def\citenum#1{{\def\@cite##1##2{##1}\cite{#1}}}
 
\newcount\@tempcntc
\def\@citex[#1]#2{\if@filesw\immediate\write\@auxout{\string\citation{#2}}\fi
  \@tempcnta\z@\@tempcntb\m@ne\def\@citea{}\@cite{\@for\@citeb:=#2\do
    {\@ifundefined
       {b@\@citeb}{\@citeo\@tempcntb\m@ne\@citea\def\@citea{,}{\bf ?}\@warning
       {Citation `\@citeb' on page \thepage \space undefined}}%
    {\setbox\z@\hbox{\global\@tempcntc0\csname b@\@citeb\endcsname\relax}%
     \ifnum\@tempcntc=\z@ \@citeo\@tempcntb\m@ne
       \@citea\def\@citea{,}\hbox{\csname b@\@citeb\endcsname}%
     \else
      \advance\@tempcntb\@ne
      \ifnum\@tempcntb=\@tempcntc
      \else\advance\@tempcntb\m@ne\@citeo
      \@tempcnta\@tempcntc\@tempcntb\@tempcntc\fi\fi}}\@citeo}{#1}}
\def\@citeo{\ifnum\@tempcnta>\@tempcntb\else\@citea\def\@citea{,}%
  \ifnum\@tempcnta=\@tempcntb\the\@tempcnta\else
  {\advance\@tempcnta\@ne\ifnum\@tempcnta=\@tempcntb \else\def\@citea{--}\fi
    \advance\@tempcnta\m@ne\the\@tempcnta\@citea\the\@tempcntb}\fi\fi}
\makeatother

\input econfmacros2.tex
\newcommand{\sla}[1]{/\!\!\!#1}
\begin{document}
\begin{titlepage}
\pubblock

\vfill
\def\thefootnote{\fnsymbol{footnote}}
\Title{Precision Higgs Physics at a \\[5pt] Future Linear Collider}
\vfill
\Author{Dave Rainwater}
\Address{\csumb}
\vfill
\begin{Abstract}
Assuming that a Higgs sector is responsible for electroweak symmetry breaking, 
we attempt to address two important questions: 
How much better precision are various measurements of Higgs boson properties at 
a future linear collider than at the LHC?
What can a future linear collider do for Higgs physics that the LHC cannot? 
\end{Abstract}
\vfill
\begin{Presented}
5th International Symposium on Radiative Corrections \\ 
(RADCOR--2000) \\[4pt]
Carmel CA, USA, 11--15 September, 2000
\end{Presented}
\vfill
\end{titlepage}
\def\thefootnote{\arabic{footnote}}
\setcounter{footnote}{0}

\section{Introduction}

The origin of electroweak symmetry breaking (EWSB) and fermion mass generation 
remains one of the most pertinent in the field of high energy physics today. 
Although there exist several explanations for EWSB, some dynamical and others 
spontaneous, preeminent among those is the existence of a Higgs sector, either 
one or two complex scalar doublets which acquire a vacuum expectation value 
(vev), or vevs, providing longitudinal degrees of freedom for the weak gauge 
bosons and additional physical observable Higgs boson states. For a single Higgs 
doublet, as in the Standard Model (SM), one neutral CP even Higgs ($H$) would be 
observed. In a two-Higgs doublet model (2HDM), such as the MSSM, five physical 
Higgses would exist: light and heavy CP even neutral scalars ($h,H$), a CP odd 
scalar ($A$), and a charged Higgs and its conjugate $H^\pm$. Additionally, a set 
of Yukawa couplings of the Higgs to SM fermions ($Y_f$), of unknown origin, 
generates the fermion masses.

No physical Higgs boson has yet been observed, although electroweak precision 
data has long suggested that the Higgs is light, of order 100~GeV. As such, it 
can possibly be accessed by present experiments, i.e. CERN's LEP II or the 
Fermilab Tevatron II. If neither of those experiments turns up evidence for a 
Higgs, the community will turn to the LHC, which will have the capability to 
detect a SM or at least one MSSM Higgs of any mass up to the unitarity limit. 

Let us suppose that the LHC finds a Higgs candidate resonance, which could be 
either the solitary physical Higgs of a single Higgs doublet model, or a 2HDM 
neutral state. This could be confirmation of previous observation at LEP or 
Tevatron, or an LHC discovery. Either way, a narrow resonance $\phi$ is found 
in one or more anticipated channels with rate commensurate with expectations. 
The task is then to determine the quantum numbers of $\phi$, first to confirm 
that it is a Higgs of some flavor, second to determine what model the Higgs 
belongs to. These quantum numbers are, with the expected value for a Higgs in 
brackets:
\begin{itemize}
\item[$\bullet$] charge [neutral]
\item[$\bullet$] color [none]
\item[$\bullet$] mass [$\O$(100)~GeV]
\item[$\bullet$] spin [0]
\item[$\bullet$] couplings (gauge, Yukawa) [model dependent]
\item[$\bullet$] total width
\item[$\bullet$] self-coupling ($\lambda$) [model dependent]
\item[$\bullet$] CP [even, odd, mixture?]
\end{itemize}

The nature of the final state of the observed channel(s) gives us at least the 
first two quantum numbers, charge and color, immediately. For example, detecting 
a Jacobian peak in the dilepton-missing transverse momentum spectrum in 
$\ell^+ \ell^- jj \sla{p}_T$ events, as expected in weak boson fusion and decay 
to a pair of $W$ bosons, would indicate that the state is neutral and colorless. 
Likewise for finding a resonance in the two photon invariant mass spectrum, i.e. 
due to Higgs production, $gg\to\phi\to\gamma\gamma$. The latter process would 
further imply by Yang's Theorem that the state cannot be spin 1, also consistent 
with a Higgs boson. A fairly precise measurement of the state's mass would also 
be obtained.

These determinations are necessary, but not sufficient conditions for confirming 
the Higgs nature of an observed resonance. To go further we must measure the 
state's coupling to weak bosons, which must be a gauge coupling, perhaps 
modified by a mass mixing parameter $\alpha$ and ratio of vevs in a 2HDM 
$\tan\beta$; and its couplings to fermions, which must be Yukawa, i.e. 
proportional to the fermion mass. Were these couplings found to meet the 
requirements of a Higgs sector, it is likely that most members of the community 
would agree that a Higgs had been discovered. However, the issue of measuring a 
self-coupling clouds this. One could argue that this is merely an aspect of 
determining the exact model that is realized in nature. To this end we would 
also want to know the CP state of the resonance.

For a SM or MSSM Higgs, the LHC can, in fact, make quite good determinations of 
the mass and gauge coupling, a very good measurement of the total width (even 
for a Higgs width smaller than the width resolution of the detectors), and a 
good measurement of at least one Yukawa coupling, $Y_\tau$~\cite{KNRZ}. 
CP-dependent distributions in Higgs production are known~\cite{gunion}, however 
observing them appears not to be possible in practice, due to detector 
effects~\cite{ATLAS}. A future International Linear Collider (ILC) can certainly 
do better than the LHC for the case of a light Higgs, approximately 
$110 \leq M_H \leq 200$~GeV. How much better an ILC could measure these 
couplings, what additional couplings it would have access to, and its ability to 
go significantly beyond LHC physics by measuring a self-coupling $\lambda$ or 
the CP nature of the Higgs sector are what I review here.

\section{The Easy Quantum Numbers}

Of greatest interest in Higgs physics is ``Where is the Higgs?'' That is, what 
is its mass? Precision fits to electroweak data suggest~\cite{EWfits} it is very 
near the lower bound from experiment, $M_H > 113$~GeV~\cite{LEP2higgs}. 
If a Higgs is found in the range $M_H < 135$~GeV or so, then the MSSM is still 
a viable theory. Finding a much heavier Higgs would suggest a different form of 
new physics. At the LHC, $M_H$ would be determined principally by observing the 
process $gg\to H\to ZZ\to 4\ell$, which is accessible at all masses. For 
example, at ATLAS for $M_H < 400$~GeV, this would yield a $0.1\%$ or better mass 
measurement, depending ultimately on uncertainty in the calorimeter 
calibration~\cite{ATLAS}.

\begin{figure}[htb]
\begin{center}
\epsfig{file=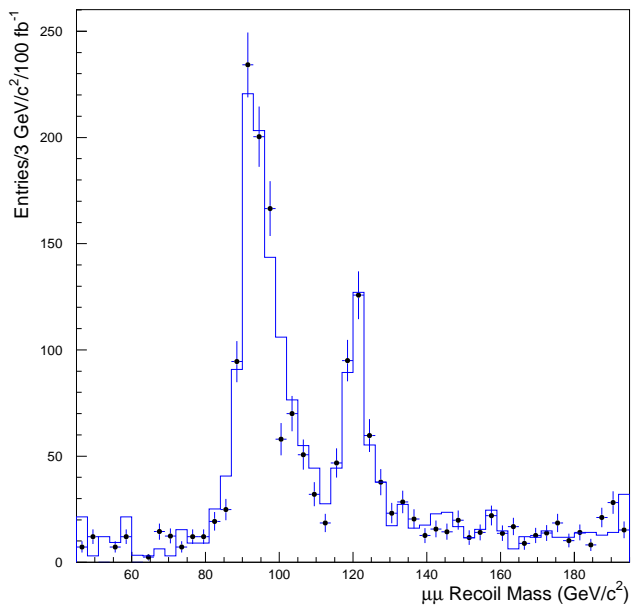,height=3in}
\caption[0]{$\mu\mu$ recoil mass for $e^+e^-\to ZH\to \mu\mu + X$,
taken from Ref.~\protect\cite{battaglia}.}
\label{fig:recoil}
\end{center}
\end{figure}

\begin{figure}[htb]
\begin{center}
\epsfig{file=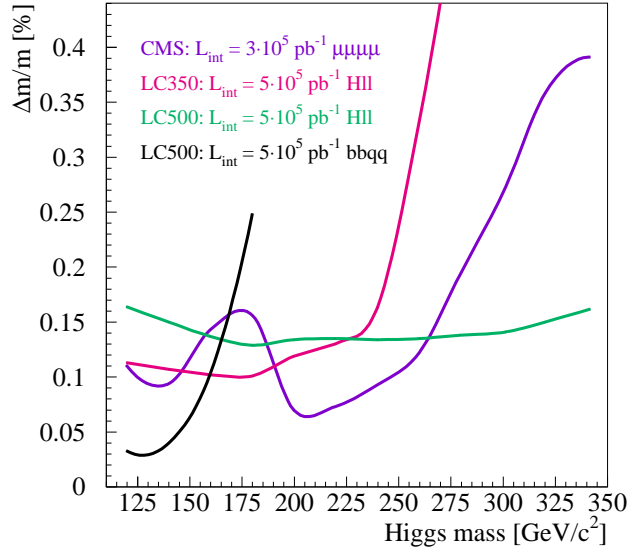,height=3in}
\caption[0]{The expected Higgs mass resolution at CMS and at 
three linear collider options, taken from Ref.~\protect\cite{sopczak}.}
\label{fig:reso}
\end{center}
\end{figure}

For a light Higgs, an ILC would improve this roughly by a factor of two to 
three, around $0.03-0.05\%$ uncertainty for a light Higgs, using a combination 
of data from the recoil mass spectrum in $e^+e^-\to HZ\to \ell^+\ell^- + X$ and 
direct reconstruction of Higgs decay into dijets. A simulated recoil mass 
spectrum for $M_H = 120$~GeV is shown in Fig.~\ref{fig:recoil}. 
Fig.~\ref{fig:reso} shows a comparison between CMS expectations and some 
different ILC energies and luminosities. However, this level of precision may be 
overkill. It would correspond to, for example, 4-loop radiative corrections in 
the MSSM! These are currently far from achievable. Additionally, the leading 
uncertainty in $\Delta M_H$ in this case is due to the large uncertainty in the 
top quark mass, $m_t$, which will not improve enough in the LHC or conceivable 
ILC experiments to warrant calculation of MSSM 4-loop corrections to $M_H$, 
possibly even the 3-loop contributions.

Determining consistency of a resonance with spin-0 has been studied thoroughly 
for the LHC. There exist two methods: angular distributions in 
$H\to\gamma\gamma$, and also in the reconstructed $Z$ bosons in 
$H\to ZZ$!\cite{ATLAS}. The LHC has no difficulty with this consistency check 
for $M_H < 400$~GeV, so I do not discuss it further here.

Measurement of the resonance's couplings to the weak bosons, photon, gluons, 
and fermions is of much greater interest. Confirming that $g_{HWW}, g_{HZZ}$ 
are gauge couplings and related by SU(2) is one of the key determinations to 
identifying the resonance as a Higgs boson. These couplings may be modified by 
mixing or other parameters of a 2HDM, which are well defined and appear only as 
overall factors. For example, the $hWW,HWW$ verticies are modified at tree 
level in the MSSM by $\sin(\beta-\alpha),\cos(\beta-\alpha)$, respectively. 
In this same vein, we would require that $g_{H\gamma\gamma}, g_{Hgg}$ are 
demonstrated to be loop-induced. 

At the LHC, $g_{HWW}$ may be determined to better than $5\%$ for a light Higgs, 
$M_H < 200$~GeV, for scenarios where the Higgs is relatively 
SM-like~\cite{KNRZ}; non-SM-like Higgs coupling extraction scenarios are just 
now beginning to be examined~\cite{mssmcoup}. This is achieved by combining 
information from various decay channels in weak boson fusion Higgs production. 
This same method also allows the total width to be extracted indirectly, to 
about the $10-15\%$ level. The best Yukawa coupling measurement can be made for 
taus, which is quite good at about $5-15\%$ over the mass region 
$115 < M_H < 150$~GeV, and work is progressing on $H\to b\bar{b}$ but this 
measurement is not likely to be better than about $30\%$ in the end~\cite{WHjj}. 
Decays to $c\bar{c}$ are completely inaccessible at the LHC, but the width to 
gluons could be determined to about $20\%$ from the rate for $gg\to H\to W^+W^-$ 
and the highly accurately known BR($WW$). These results would already be quite 
good and allow for considerable model determination, but an ILC could do far 
better.

\begin{figure}[htb]
\begin{center}
\epsfig{file=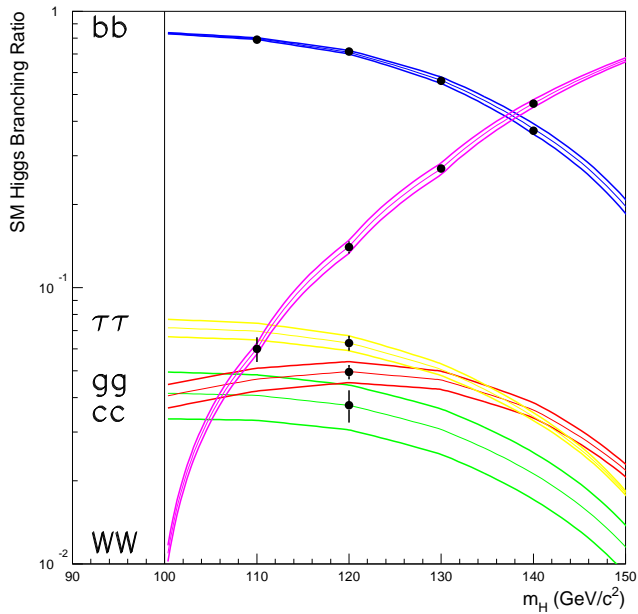,height=3in}
\caption[0]{Estimated accuracy for SM Higgs branching fractions for 
500~fb$^{-1}$ of data at TESLA, taken from Ref.~\protect\cite{battaglia}.}
\label{fig:br_sm}
\end{center}
\end{figure}

The procedure for extracting couplings\footnote{We may alternatively discuss the 
measurement of partial widths, which are directly proportional to the coupling 
squared. If the total width cannot be determined, it is more appropriate to 
discuss measurement of branching ratios.} is equally involved at the ILC. First, 
the recoil mass spectrum in $ZH$ production would be used to determine 
$\sigma_{ZH}$ to about $2\%$ (TESLA) for a light Higgs, say 
$110 < M_H < 150$~GeV. This measurement is the basis for extracting absolute 
branching ratios. Once the $WW$ branching ratio is known ($< 5\%$ for a light 
Higgs, possibly as good as $2\%$), $g_{HWW}$ could be obtained to better than 
$2\%$. From there, Yukawa couplings could be extracted directly. The expected 
uncertainty in the branching ratio for decay to $b\bar{b}$ is anticipated to be 
the lowest, about $2\%$ for $M_H = 120$ with reasonable luminosity; $8\%$ for 
$c\bar{c}$ and $6\%$ for $\tau^+\tau^-$. Additionally, BR($gg$) could be 
determined to about $8\%$ for the same mass. Fig.~\ref{fig:br_sm} shows the 
probable branching ratio measurement precision at TESLA.

$t\bar{t}H$ production is possible for a high-energy ILC, being optimum at about 
$\sqrt{s} = 800$~GeV for $M_H = 120$~GeV. There is considerable discrepancy in 
the literature, however, as to the achievable precision in 
$\delta g_{Htt} / g_{Htt}$, ranging from $6\%$~\cite{ttH_1} on the optimistic 
end to as pessimistic as about $50\%$~\cite{ttH_2}. Clearly, more study is 
needed to resolve this disagreement. Finally, the total width could be had to 
better than $5\%$ for $M_H = 120$~GeV, or about $3\%$ for $M_H = 140$~GeV, using 
the $H\nu\nu$ cross section as key input, which contains the already determined 
gauge coupling.

\begin{figure}[htb]
\begin{center}
\epsfig{file=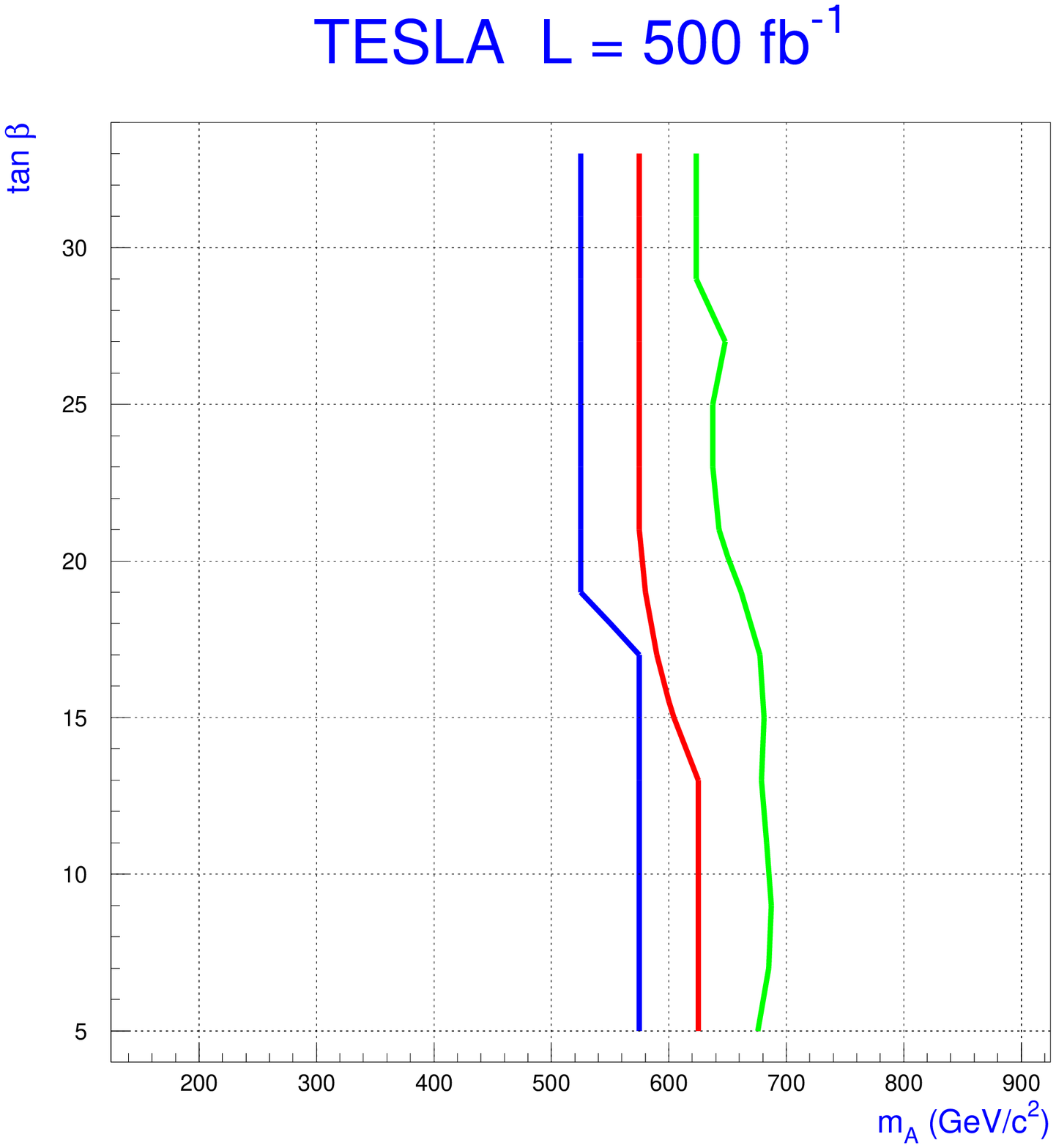,height=3in}
\caption[0]{SM v. MSSM discriminating power as a function of $M_A$ and 
$\tan\beta$ for 500~fb$^{-1}$ of data at TESLA, 
taken from Ref.~\protect\cite{battaglia}. Parameter space to the right of the 
curves would appear SM-like. The blue curve is $95\%$ CL, red is $90\%$ CL, 
and green is $68\%$ CL.}
\label{fig:discrim}
\end{center}
\end{figure}

As far as distinguishing the SM from the MSSM, prospects appear good at the ILC 
but this subject begs for more study. Ratios of branching ratios are important, 
especially BR($b\bar{b}$)/BR($W^+W^-$), which could yield an indirect 
measurement of $M_A$. (It is possible that observing the CP odd state $A$ itself 
could prove difficult.) Other important ratios are BR($c\bar{c}$)/BR($b\bar{b}$) 
and BR($gg$)/BR($b\bar{b}$). It is known that an MSSM Higgs sector can be 
established as non-SM-like at the $95\%$ CL for $M_A < 550$ or so at TESLA, 
depending somewhat on $\tan\beta$. This is shown in Fig.~\ref{fig:discrim} for 
500~fb$^{-1}$ of data.

For a Higgs heavier than about 150~GeV, measurement of Yukawa couplings (other 
than to the top quark) are essentially inaccessible at any machine, except in 
certain restricted regions of a 2HDM. However, it is still highly desirable to 
know the gauge couplings and total width in this regions. Preliminary results 
indicate~\cite{FNAL_LC} that for the mass region $150 < M_H < 300$~GeV an ILC 
could achieve $10\%$ uncertainty for BR($WW$) and $10-25\%$ uncertainty for 
BR($ZZ$), the latter depending strongly on $M_H$. For $M_H > 300$~GeV, BR($VV$) 
could be determined to a few percent, and BR($t\bar{t}$) becomes accessible, 
probably at the $6\%$ level. It is currently not known how to measure this at 
all at the LHC. For $M_H > 230$~GeV or so, the Higgs width is expected to 
exceed detector resolution; the LHC and an ILC would have comparable ability 
here, at the few percent level for a heavy Higgs. A heavy Higgs is widely 
regarded as less interesting, and as such this scenario has not received as much 
attention in studies for an ILC.

It is easy to find already in the SM that radiative corrections are important. 
For example, $\delta$~BR($b\bar{b}$)$> 30\%$ for $M_H = 120$~GeV, simply due to 
the gluonic corrections. In extensions to the SM, such as the MSSM, large 
radiative corrections to the couplings (partial widths) lurk behind every 
corner. To take a couple examples, light squark corrections to $h\to gg$ can be 
as large as factors of 2-3 in the partial width, but disappear as the squarks 
get heavy, greater than a few hundred GeV. Also, gluino or chargino corrections 
to the $b$-$H$ verticies and $H_1,H_2$ mixing can lead to order 1 corrections to 
the partial widths to $b$ quarks or $\tau$ leptons~\cite{carena}. Considerable 
effort has been invested in calculating such corrections over the past decade, 
but the state of the art advances and interesting phenomenology continues to be 
revealed.

\section{The Tough Quantum Numbers}

While the ``easy-to-determine'' quantum numbers of a newly discovered resonance 
may be sufficient to establish it as part of a Higgs sector, in some sense the 
more difficult measurements are the interesting ones. These are the CP nature of 
the resonance, more specifically if there is any CP mixing in the neutral states 
in the case of a two Higgs doublet sector; and the self-coupling(s) $\lambda_i$. 
In the SM $\lambda$ is related to the Higgs mass via $M_H^2 = 2 v^2 \lambda$, 
and as it is a free parameter, $M_H$ is undetermined. In the MSSM, the various 
$\lambda_i$ are gauge couplings, thus $M_H$ is constrained. To be convinced 
that an observed Higgs sector (and perhaps other MSSM candidate states) belong 
to the MSSM, we would need to verify that these are, indeed, gauge couplings. 
Thus, multiple Higgs production would have to be observed.

Several studies have addressed the issue of multiple Higgs production at an ILC, 
highlighting scenarios where the cross section is large enough to obtain a 
substantial rate~\cite{hhh}. Only two groups have performed a signal v. 
background study at the phenomenological level (in the $ZHH$ channel for the 
SM), and only one of those two groups, Castanier et al., included detector 
simulation. However, their results are quite promising. For a light SM Higgs, 
i.e. $120 < M_H < 140$~GeV, their study suggests that the $ZHH$ cross section 
could potentially be measured at the $12-18\%$ level for large integrated 
luminosity, 2000~fb$^{-1}$. They further show that this would translate to about 
a $20\%$ measurement of $\lambda_{hhh}$. It is clear that studies such as these 
at a future linear collider will depend critically on $b$-tagging performance. 
The studies go on to point out that in some MSSM parameter space regions, heavy 
Higgs decay to lighter Higgs states has significantly larger rates than the SM 
case, making the prospects for observation quite good. In contrst, it remains 
to be shown that the LHC has any capability to measure a Higgs sector 
self-coupling.

Research into methods to extract the CP nature of observed neutral Higgses is 
even less developed. Studies so far indicate that a CP = +1 state can be 
qualitatively distinguished from a CP = -1 state via angular distributions in 
$ZH$ production~\cite{CP}, but if the state has mixed CP, the -1 component is 
very easily washed out. Additional work is sorely needed in this area, as the 
LHC again has no capability here, primarily due to the strong backgrounds and 
detector effects that hide the relevant distributions.

\section{Conclusions}

Prospects for observing a Higgs sector as the responsible mechanism for EWSB are 
quite good: precision fits to electroweak data suggest that the Higgs is light 
and thus accessible, perhaps by the Tevatron, and the LHC has the capability to 
observe a Higgs boson of any mass up to the unitarity limit. While the Tevatron 
could not make any serious measurements of couplings or other important 
properties of an observed resonance to completely convince one that it is part 
of a Higgs sector, the LHC can go a long way toward this goal. The LHC would 
determine the mass to a greater precision than can be matched theoretically, and 
would be able to determine the gauge coupling and total width of a Higgs boson, 
either directly or indirectly, to better than the $10\%$ level. However, the LHC 
will have considerable difficulty to observe most fermionic decays, and will not 
have the capability to observe any self-couplings or identify CP mixing among 
the states of multiple physical Higgses.

An ILC extends our knowledge of a Higgs sector considerably. Its measurements of 
both gauge and Yukawa couplings would be superior over most of the mass range of 
a possible Higgs, and it would have access to additional fermionic decays as 
well as self-couplings and probably CP mixing, improving model discrimination. 
However, the state of the art on the last goal is still somewhat underdeveloped. 
Also lacking is a detailed overview of the capability of an ILC to distinguish 
different models based on coupling and mass measurements, although at least one 
study to address this is nearing completion~\cite{LCcoup}.  So far, studies have 
presented levels of measurement precision based only on cross sections in the 
SM. If a Higgs sector turns out to be not very SM-like, then some of these 
levels of precision could be quite poor, while others are better than expected, 
and the restriction in model parameter space may or may not be satisfyingly 
small.

\Acknowledgments
I am grateful to the organizers of the conference and to the many physicists 
who have performed the LHC and LC Higgs studies which I have attempted to 
summarize here.

\end{document}

%% file: econfmacros2.tex
\def\beq{\begin{equation}}
\def\eeq#1{\label{#1}\end{equation}}
\def\eeqn{\end{equation}}

\newenvironment{Eqnarray}%
   {\arraycolsep 0.14em\begin{eqnarray}}{\end{eqnarray}}
\def\beqa{\begin{Eqnarray}}
\def\eeqa#1{\label{#1}\end{Eqnarray}}
\def\eeqan{\end{Eqnarray}}

\let\bar=\overbar

\def\O{{\cal O}}

\def\Dslash{\not{\hbox{\kern-4pt $D$}}}
\def\dslash{\not{\hbox{\kern-2pt $\del$}}}

\def\msb{{\bar{\ssstyle M \kern -1pt S}}}

\def\lsim{\mathrel{\raise.3ex\hbox{$<$\kern-.75em\lower1ex\hbox{$\sim$}}}}
\def\gsim{\mathrel{\raise.3ex\hbox{$>$\kern-.75em\lower1ex\hbox{$\sim$}}}}